\documentclass[onecolumn]{elsart3p}

\usepackage{amssymb}
\usepackage{amscd}
\usepackage{makeidx}
\usepackage{graphicx}
\usepackage{amsmath}
\usepackage{epsfig}
\usepackage[dvips]{color}

\DeclareMathOperator{\N}{\mathbb{N}}

\DeclareMathOperator{\Z}{\mathbb{Z}}

\newcommand{\beq}{\begin{equation}}
\newcommand{\enq}{\end{equation}}
\newcommand{\es}{\end{subequations}}

\newcommand{\la}{\lambda}

\newcommand{\TT}{\mathbb{T}}
\newcommand{\uu}{\mathbf 1}

\def\Journal#1#2#3#4#5#6{#1. #3. #4, #2; #5: #6.}
\def\Book#1#2#3#4#5{#1. #3. #4: #5; #2.}

\def\a{\alpha}

\def\be{\beta}

\def\d{\delta}

\def\e{\epsilon}

\def\eq{\equiv}

\def\f{\frac}\def\fo{\forall}
\def\g{\gamma}

\def\l{l}

\def\o{\omega}

\def\un{\underset}

\begin{document}

\begin{frontmatter}
\title{On properties of Continuous-Time Random Walks  with
Non-Poissonian  jump-times}
\author{Javier Villarroel\corauthref{cor}}
\ead{javier@usal.es}
\address{Facultad de Ciencias, Universidad de Salamanca.
Plaza Merced s/n, E-37008 Salamanca, Spain}
\corauth[cor]{Corresponding author. Address: Facultad de Ciencias, Universidad de Salamanca.
Plaza Merced s/n, E-37008 Salamanca, Spain. Fax: (+34) 923 294514.}
\author{Miquel Montero}
\ead{miquel.montero@ub.edu}
\address{Departament de F\'{\i}sica Fonamental, Universitat de
Barcelona. Diagonal 647, E-08028 Barcelona, Spain}
\maketitle

\begin{abstract}

The usual development of the continuous-time random walk (CTRW)
 proceeds by assuming  that  the present  is  one of the
jumping times. Under this restrictive assumption   integral
equations for the  propagator  and mean escape times have been
derived. We  generalize these  results  to the case when the
present is an arbitrary time by recourse to renewal theory. The
case of Erlang distributed times is analyzed in detail. Several
concrete examples are considered.

\end{abstract}

\end{frontmatter}
\section{Introduction}

In this article we study   transition probabilities and    mean
exit times of continuous-time random walks (CTRWs). By this    we
understand a   random process $X(t)$ whose evolution occurs purely
 via  jumps of a random magnitude $\Delta  X(t)$ that happen at \it random times \rm $t_n, n\in \Z$     where   the
 ``waiting times" $\Delta t_n \equiv t_n-t_{n-1}$ are independent
 and
 identically distributed (iid).
CTRWs have revealed as an interesting tool to model a large
variety of physical phenomena that undergo sudden
  random  changes. From
a Mathematical perspective the consideration and interest in
those processes  can be traced back to  the seminal work of
Kolmogorov
  \cite{ko} and Feller   \cite{f40}. Applications to describe changes of stock
markets due to unexpected catastrophes  were first noted in  the
seminal work of Merton~\cite{m76}, where it  is assumed that
inter-catastrophe times are exponentially distributed independent
of the magnitude of the catastrophe, i.e.,  that catastrophes are
driven by a compound Poisson process (CPP). CTRWs generalize in
an important way   the latter processes allowing for general
distribution of the waiting times. Correlation between waiting
times and jumps is also permitted.

In Statistical physics CTRWs  became popular after the work of
Montroll $\&$ Weiss \cite{mw65}, and have been used to describe
physical phenomena ever since.  To list a few examples we note
applications  to earthquake
 modelling (e.g., \cite{vj78,hs02}), rainfall description  \cite{rici88} or to transport in disordered media (e.g.,  \cite{ms84}). More recently,
 the use of CTRWs has been advocated to give a microscopic, tick-by-tick, description of financial markets: see
  \cite{s07,ks03,mmw03,mmp05,mmpw06,mpmlmm05}.

Unfortunately,   the Markovian nature of  CPP does not extend to
the more general CTRW.   In spite of this ominous situation,
CTRWs satisfy a pseudo-Markovian property, namely that the
knowledge of the \it past-prior to  the last jump \rm provides no
further information    to   determine the future evolution than
merely knowing the state of the system  at such a jump time
---cf. Eq.~(\ref{memoryless}).
   This fact  explains why   scholars  have usually  focused   in
   studying
 the statistics of CTRWs \it right after a jump occurs\rm.  In this regard, a linear integral
equation for the mean escape times off a given interval has been
derived \cite{mmp05,mmpw06}.
Similarly, the basic probability for the process to
  be found in a certain  region,   given the position at a jump time,  satisfies a certain     integral equation  first derived by Weiss
  \cite{w94,mw65}.

However,  this setting |wherein the present  must be  one of the
jumping times| \it does not cover the most general situation\rm.
The relevance of this fact  is further stressed by  noting  that
in several physical problems one may not be able even to decide if
such arrival has occurred. Hence, the   issue of how to
generalize the aforementioned framework to \it arbitrary present
times
 \rm arises naturally.
In this sense some extensions have been considered in the past
\cite{w94,JT76}, where it is assumed that the process {\it
starts\/} at $t=0$ and behaves in a different way in the time
period prior to the first jump.
  Nevertheless, to our knowledge, there is no robust and self-contained development on this topic in the literature.
 This paper tries to  fill this gap  and considers, in particular, the determination of the propagator. We also show how to obtain the mean exit time.
  We  find that, unless jumping times are exponentially distributed,  the results of \cite{w94,mmp05} must be
  corrected.  The  implications of this fact to the calculation of  the    correlation
  function of the process are obvious. We consider here the simpler uncorrelated case: the  more general case when jumps and sojourn times are correlated requires new ideas and
  will be the subject of a future publication.

  The structure of the paper is the following. In
section~\ref{s2} we obtain  the unrestricted propagator of a CTRW
and relate it
 with  that corresponding to    starting at jump times. This
connection  involves     a certain
  object  whose distribution    may be found by recourse to classical renewal
theory\cite{c65,kt81}.    General correlation functions follow
immediately.
 Section~\ref{s3} is devoted to solving the integral equation for the after-jump propagator by means of a joint Fourier-Laplace
 transform.    In section~\ref{s4} we extend the results of Masoliver et
al.~\cite{mmp05}  and obtain the  mean escape time  off intervals
\it starting at an arbitrary instant\rm.     In section~\ref{s5},
concrete, explicit formulae are given
 in the case that  waiting times have  Erlang distribution.   Recently, Erlang times have been the
subject of much interest in the context of information traffic and
phone-calls waiting times \cite{sb00,fc02}.   In the context of
transaction orders in financial markets    the appearance of this
distribution  can also be expected since
 it takes, at least, two arrivals (buy and sell orders) for  a transaction to be completed.   For further applications to ruin
problems and insurance see  \cite{dh01}.

To help a reader not familiar with measure theory     we assume that  all distributions involved have
   a density.  However, we find that all results    extend  to
  a general situation.

\section{Fundamentals of CTRWs} \label{s2}   
\begin{figure}[hbtp]
{\hfil\includegraphics[width=0.90\textwidth,keepaspectratio=true]{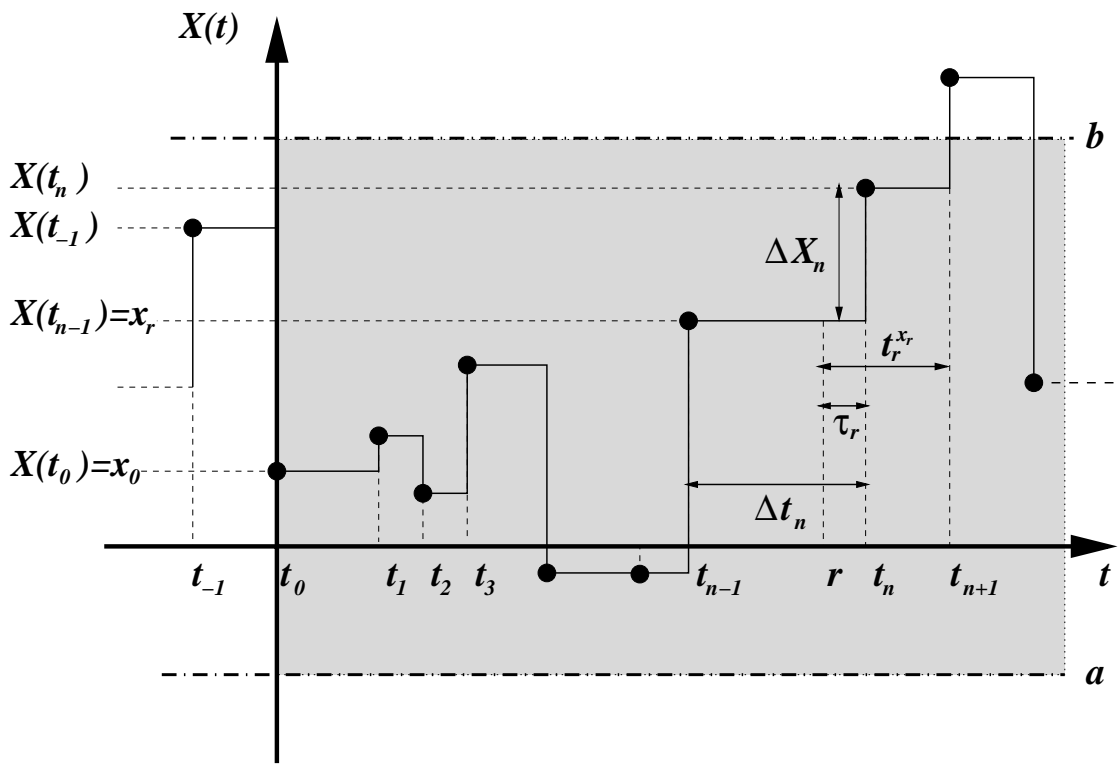}}
\caption{A sample path of the process $X(t)$.} \label{Fig1}
\end{figure}

Recall that any realization of   a  CTRW is  given by a series of step functions in such
a way that $X(t)$ changes at random times $\dots, t_{-1}, t_0,
t_1, t_2,\dots, t_n, \dots$, while it remains fixed in place
between successive steps |see Fig.~\ref{Fig1}. The interval
between these successive steps defines a sequence of independent
identically distributed (iid)  random variables $\Delta
t_n=t_n-t_{n-1}$, the waiting times. The (random) change, or
jump, at $t_n$ in the process is given by $\Delta
X_n=X(t_n)-X(t_{n-1}) $. We assume that  $\vec Z_n\eq (\Delta
X_n, \Delta t_n)$ defines a sequence of iid two-dimensional
random variables  \it having \rm     joint and marginal
densities  $\rho(x,\tau)$,  $h(x)$ and $\psi(\tau)$ given by\
\begin{equation}
\rho(x,\tau)dxd\tau =\Pr\big(x\le \Delta X_n\leq x+dx, \tau\le \Delta
t_n\leq \tau+d\tau\big),
\end{equation}
\begin{equation}
h(x)dx=\Pr \big(x\le\Delta X_n\leq x+dx\big),\
\psi(\tau)d\tau=\Pr\big(\tau\le\Delta t_n\leq \tau+d\tau\big). \label{h}
\end{equation}

 \subsection{The propagator and related quantities}
  One of the  basic objectives
within the  theory of stochastic process is to obtain the future evolution given
the actual state of the system; in this regard the main object is
the conditional density of $X(t)$ given the  present position
$X(r)\eq x_r, r\leq t$ or, in more physical terms,  the
propagator $p(x, t|x_r,r)$.   If $t=r+\tau$ where  $\tau\geq 0$
the propagator reads
\begin{equation} p(x,r+\tau|x_r,r)dx=\Pr \big(x\leq X(r+\tau)\leq
x+dx|X(r)=x_r\big). \label{propagator}
\end{equation}
  Note that
$r$, $t$ and every $t_n$ are clock times; this is  unlike $\tau $
which  gives the measure of a time interval.  In order to have a
frame of reference, we must specify some time origin. For
convenience we    take it to coincide with one of the
 jump times,   $t_0=0$  say. In some problems $t_0$ may be
identified as the starting point of the process $X(t)$, but this
is just one possible occurrence |see Fig.~\ref{Fig1}.

Now, recall that a generic CTRW  is  non-Markovian, and hence
knowledge of the past gives additional information to that already
provided by the state of the system at present.  An exception to
this occurs when  the present  happens to be one  of the jump
times:
 transition probabilities \it starting at a jumping
 time \rm
$t_n $,  show a pseudo-Markovian property:
\begin{eqnarray}
\Pr \big(x\leq X(t_n+\tau)\leq
x+dx|X(t_n)=x_n, X(t_{n-1})  = x_{n-1},\dots 
\big)=  \nonumber \\
\Pr \big(x\leq X(t_n+\tau)\leq x+dx|X(t_n)=x_n\big)\eq
\Pi(x,t_n+\tau|x_n,t_n)dx, \label{memoryless}
\end{eqnarray}
and, whenever $r\le t_n$ is
 \beq \Pr \big(x\leq X(t_n+\tau)\leq x+dx|X(t_n)=x_n, X(  r)=x_r  \big)
 = \Pi(x,t_n+\tau|x_n,t_n)dx,\label{memoryless2}\enq
which is the backbone of these {\it renewal\/} processes.
Intuitively, once a jump happens the system  \it forgets all the
past   previous   to the jump time  starting anew\rm. It must be
stressed that $\Pi(x,t_n+\tau|x_n,t_n)$ in the expression above
could be interpreted as the propagator starting from the jumping
time $t_n$ |i.e., roughly $\Pi(x,t_n+\tau|x_n,t_n)$ is \it
interpreted \rm as $p(x,t_n+\tau|x_n,t_n)$ whereas when we write
$p(x,r+\tau|x_r,r)$ the time $r$ is arbitrary. Further, it can be
proven that $\Pi(x,t_n+\tau|x_n,t_n)$ \it must \rm satisfy the
homogeneity condition \beq \Pi(x,t_n+\tau|x_n,t_n)=\Pi(x-x_n,
\tau|0,0)\eq \Pi(x-x_n, \tau ). \enq
 These properties   imply that $\Pi(x,t )$ solves  the following integral
equation (see   Weiss~\cite{w94}): \beq  \Pi(x,t )=
\delta(x)\big(1-\Psi(t)\big)+\int_0^{t }\int^{+\infty}_{-\infty}
\rho(x',t') \Pi(x-x',t-t')dx' dt', \label{Pi_int} \enq where
$\Psi( t) \eq \Pr\big(
 \Delta t_n\le  t\big)$.   Note that a class of CTRWs for which this
equation can be solved in a closed form has been given in~\cite{mmpw06}.  We elaborate on this in
 section~\ref{s3} below.

  However,    although the assumption that the present is
one of the jumping times  may be convenient, there is,
unfortunately, no convincing reason to that effect,    neither
from a mathematical nor a physical point of view.     The
evolution of the system is then described by  the more general
object $p(x,r+\tau|x_r,r)$ and the issue of determining it arises
  in a natural way. Here we
 address this problem, assuming for the sake of simplicity, that   jump
times and jump-magnitudes, $\Delta t_n$ and $\Delta X_n$, are
mutually independent, i.e., that
 $\rho(x,t)=\psi(t)h(x)$. The general, correlated case will be the
 subject of a future publication.

To this end set in Eq.~(\ref{propagator})
 $ p=p^{(0)}+p^{(1)}$ where   $p^{(0)}(  B,r+\tau|x_r,r )$  is the
probability of a transition from $x_r$ to $B\eq [x,x+dx]$  in the
time interval $(r,r+\tau]$ occurring with no jumps.  Obviously
 \begin{eqnarray*}
p^{(0)}(  B,r+\tau|x_r,r )dx\eq \Pr \Big( X_{r+\tau}\in
B,N_{(r,r+\tau]}=0\Big|X_r=x_r\Big) 
=  \d(x-x_r)\Pr \big(
 N_{(r,r+\tau]}=0\big)dx.
\end{eqnarray*}
Here
 $N_{(r,r+\tau]}$ denotes  the number of jumps of $X $ on
 $(r,r+\tau]$.
We next evaluate   the probability  that a such  transition occurs
with   one or more jumps:
\beq  p^{(1)}(  B,r+\tau |r,x)dx\eq
\Pr\Big( X_{r+\tau}\in B,N_{(r,r+\tau ]} \ge 1 \big |X_r=x_r\Big).
\enq     To this end we introduce the
``excess life" $\tau_r$ as follows:      $  r+ \tau_r $  is the
  time
 at which the first jump past $r$ occurs.
Note that $\tau_r$ is random, in contrast with
   $\tau$ above, which is a number.  Obviously, $ r+ \tau_r = t_n$ for
some $n$  (see Fig.~\ref{Fig1} above).  By conditioning respect
 to this object one  has, by the total probability theorem  and the
memoryless property~(\ref{memoryless}) \beq   p^{(1)}(  B,r+\tau
|x_r,r) =  $$ $$ \int_r^{r+\tau}\int^{+\infty}_{-\infty}
  \Pr \Big(   X_{r+\tau}\in B,N_{(r,r+\tau ]}\ge 1\Big| X_{t_n }=x',t_n
=t'\Big) \Pr \Big(X_{t_n }\in dx', t_n
  \in dt'\Big|X_r=x_r\Big). \enq
Note that we use the convenient notation $ \Pr \big(   X_{t_n }\in
 dx'\big)\eq  \Pr \big(   X(t_n)\in [x',x'+dx']\big)$ and so forth.  A further simplification arises from    the independence
 assumptions:
  \begin{eqnarray*}
&& \Pr \Big(   X_{r+\tau}\in B,N_{(r,r+\tau ]}\ge 1\Big| X_{t_n
}=x',t_n =t'
 \Big)=   \Pi(x-x', r+\tau-t')dx,\\
 && \Pr \Big(   X_{t_n }\in dx', t_n \in dt'\Big|X_r=x_r\Big)=   h(x'-x_r)dx'  \Pr \Big(      \tau_r \in
  [t'-r,t'-r+dt'] \Big).
 \end{eqnarray*} Hence,
by substitution into the RHS of~(\ref{propagator}) we find
   \begin{eqnarray}
 &&p^{ }(  x,r+\tau |x_r,r) 
 =\d(x-x_r)\Pr \big(
  \tau_r>\tau \big)+ \int_0^{\tau }\int^{+\infty}_{-\infty}  \Pi(x-x',
\tau -\tau') \Pr \Big(      \tau_r \in
 d\tau' \Big)h(x'-x_r)dx' =\nonumber \\
&& \d(x-x_r)  \big(1-
  \ \Phi(\tau|r )\big)+ \int_0^{\tau}\int^{+\infty}_{-\infty}  \Pi(x-x_r-x',
\tau -\tau')  \phi(\tau'|r)h(x')dx' d\tau'. \label{prop_int}
\end{eqnarray}
  Here $\Phi(\tau|r)$ and $ \phi(\tau|r)  $ denote accumulated distribution
function and  the density of  $\tau_{r  }$:
\begin{eqnarray}
\Phi( \tau|r)\eq  \Pr \big(   \tau_r \le   \tau \big)
=\sum_{n=1}^\infty\Pr\big(t_{n-1}\le  r< t_n\le  r+\tau\big)  \
\text{ and } \phi(\tau|r) \equiv \frac{d\Phi(\tau|r)}{d\tau}.
\label{full_p_int}
\end{eqnarray}

To summarize, the propagator is   recovered by~(\ref{prop_int})  if
$\Phi(\tau|r)$ and $\Pi(x,t)$ are known.\footnote{Two-point and
correlation functions follow  immediately from Eq.~(\ref{prop_int}).} The
latter can
   be completely retrieved by means of a joint Fourier-Laplace transform
 (see section~\ref{s3}). It turns
out that
   the former     can also be \it  recovered in closed form \rm
   by  means   of classical renewal
theory  as we now show. To this end   let $m(t)$ denote
 the  renewal function, the      mean   number
of jumps between $t_0=0$ and $t$.    It satisfies the integral
renewal
 equation (see \cite{c65} or \cite{kt81})
 $ m(t)=  \Psi(t) +\int_0^t  \Psi(t-t')dm(t')    $.
Further,
 by the law of total
probability one finds that $ \Phi(\tau|r)$ solves the following
renewal equation
 \beq\Phi(\tau|r)=   \Psi(r+\tau)+\int_0^r
\Big(\Phi(\tau|r-r')-1\Big) d\Psi(r') \label{Psi_tr},\enq   whereupon, appealing
to  the renewal's theorem   one finds that $\Phi(\tau|r), \phi(\tau|r)$
are given by \beq \Phi(\tau|r) = \Psi(r+\tau)- \int_{0}^{r }
\Big(1-\Psi(r+\tau-t')\Big)dm(t'),\ \phi(\tau|r)=\psi( r +\tau) +
\int^{r}_{0} \psi(r+\tau-t') dm(t'). \label{rep}\enq

 Laplace  transformation is useful to  evaluate in an explicit
 form  the latter objects.  To this end note that manipulation of Eq.~(\ref{rep}) yields also
\beq \Phi(\tau|r) = \int^{r+\tau }_{r}\Big(1-
\Psi(r+\tau-t')\Big)dm(t').\label{repp}\enq Let $\hat  \phi(s|r) $ and
$ \hat \psi(s)$  be  the
 Laplace transforms of
  $\phi(\tau|r)$  and $\psi(\tau)$.  Upon Laplace transformation,
  we find
  \beq  m'(t)\equiv\frac{d m(t)}{dt}=\frac{1}{2\pi i}
\int_{c-i\infty}^{c+i\infty} e^{ st}\f {\hat \psi(s)}{1-\hat
\psi(s)} ds,\   c>0,\label{def_m}\enq  \beq  \phi(\tau|r)
=\frac{1}{2\pi i} \int_{c-i\infty}^{c+i\infty} e^{ s (\tau+r)}
\Big(1-\hat \psi(s)\Big) \int_r^\infty e^{-st}dm(t) ds,\   c>0
\label{hat_phi}. \enq

 The above results recover the propagator in  closed form. We next  mention    properties of this object that follow from  Eq. (\ref{prop_int}).
 Note first
that       $p$  is spatially-homogeneous: $p(x,r+\tau |x_r,r)=
p(x-x_r,r+\tau|0,r)\equiv p(x-x_r,r+\tau|r)$.   This in turn
implies that $X_{t_2}-X_{t_1}$ is independent of $X_{t_1} $ for
any two times $t_1\le t_2$.  This does not imply, though, the
stronger property of independence of the  increments  as in
L\'{e}vy processes.

   By contrast, $p$  is homogeneous in time  only when   $ \Phi(\tau|r) $ is independent of $r$:  $ \Phi(\tau|r)
=\Phi(\tau| 0)\eq\Psi(\tau)$.    Similarly, a comparison between
(\ref{Pi_int}) and (\ref{prop_int}) shows that for the condition
$p^{ }( x,r+\tau |r) =\Pi( x,\tau)$  to hold  one needs, again,
that $ \Phi(\tau|r) =\Phi(\tau| 0)$. It can be proven that the
latter  holds only if waiting times  are exponentially
   distributed. If this is not the case  \it the after-jump propagator
  $\Pi$  does  not describe the temporal evolution \rm of the system; instead,  the more general object
  $  p(x-x_r,r+\tau|r)$ must be used.

  The non-Markovian character of the process (see section \ref{s3})   implies that any available information about
prior evolution of the process affects the transition probability.
 Suppose  for instance we wish to determine the future evolution of the system if   we know that  $X(r)= x_r$ and, in addition,  the   value
  of the  \it previous   \rm jumping time,    say $t_{n-1}$.
With no loss of generality set $t_{n-1}=t_0=0$.   The transition
probability adapted to that information is
\begin{equation}
\pi(x,r+\tau| x_r,r)
dx \equiv \Pr \big(x\leq X(r +\tau)\leq x+dx\big |X(r')= x_r,\fo r'
\leq r \big).
\end{equation}
With similar ideas to those used before one can prove that
\begin{equation}
 \pi(x,r+\tau| x_r,r)
=\pi(x-x_r,r+\tau| 0,r)=$$ $$ \delta(x-x_r)\frac{1-\Psi(r+\tau
)}{1-\Psi(r )}+\int_{0}^{\tau} \int_{-\infty}^\infty h(x')
\frac{\psi(r+\tau')}{1-\Psi(r)}\Pi(x-x_r-x',\tau-\tau')dx' d\tau'.
\end{equation}

\section{Transition probabilities in frequency domain}\label{s3}

    Since Eq. (\ref{Pi_int})    is of convolution type   an explicit analytical solution can be retrieved in
  the  Fourier-Laplace domain.  We remind how this is done |for other considerations on this regard see also \cite{w94,mmpw06,ba,go}.
 By taking  a  Fourier (Laplace) transform in space (respectively, in
time)  Eq.~(\ref{Pi_int}) yields  that the joint Fourier-Laplace
transform of $\Pi$ is recovered as:
\beq
\hat{\tilde \Pi}(\omega,s)
\eq \int_{0}^\infty {\int_{ - \infty }^{   \infty } {\Pi(x,t
)e^{i\omega x -st} dxdt}}= \frac{ 1-\hat\psi (s)}{{\big( 1 -
\tilde h(\omega) \hat \psi(s)\big) s}}. \label{hat_tilde_Pi}
\enq
Here $\o$ is a real variable while $s\eq s_R+is_I$ is complex,
  $s_R>0$ and  we  define
 \begin{equation}
\tilde h(\omega)  \eq \int_{ - \infty }^{   \infty }
h(x)e^{i\omega x} dx, \  \ \hat{\psi}(s) \eq\int_{0}^\infty
\psi(t) e^{ - s t}dt,\ \ \hat{\Psi}(s)  \eq \int_{0}^\infty
\Psi(t) e^{ - s t}dt= \hat\psi (s)  /s.
 \end{equation}

 By inversion one could, in principle,  recover both $p(x,r+\tau|r) $ and
 $\Pi(x,r)$.     Regarding the latter we note that    $\hat{\tilde \Pi}$ \it must \rm be  a combination of
an    isolated  mass at $ x=0$      and a sub-stochastic density
$\hat{\Pi}^{(1)}(x ,s)$  which is the inverse FT of $Q\eq
\hat{\tilde \Pi}( \o,s) -  \hat{\tilde \Pi}(\infty,s)$. Thus, the
inversion formula reads  \beq
   \Pi(x,t)= \d(x)\Big(1-\Psi(t) \Big)+ \frac{1}{4\pi^2 i} \int_{c-i\infty}^{c+i\infty} e^{st}    \int_{-\infty}^{+\infty} e^{ -i\o x}
   Q(
\o,s)   d\o ds, \ c>0.\label{Pi_xt}\enq

   Recall that
 for $X_t$  to be  Markovian, both the propagator $p$ and after-jump
propagator $\Pi$ must satisfy the Chapman-Kolmogorov equation.
Using the spatial invariance, we find that
 this equation   reads, in Fourier
 domain, as
$\tilde\Pi(\o,t+l)= \tilde\Pi(\o,t) \tilde\Pi(\o,l)$ for all $\o$
and times $t,l$. This is   Cauchy's functional whose solution is
$\tilde\Pi(\o,t) =A(\o)e^{-B t}.$       Taking a  further Laplace
transformation (in time) a comparison   with
Eq.~(\ref{hat_tilde_Pi}) shows that  we must require  $ A(\o)=1,
 B(\o)=\lambda\big(1-\tilde h(\o)\big) \text { and  } \hat
\psi(s)=\lambda/(\lambda+s)  $   for some positive $\lambda$, i.e.
that times must be exponentially distributed for the process to
be Markov.

\begin{figure}[hbtp]
\begin{center}
\includegraphics[width=1\textwidth,keepaspectratio=true]{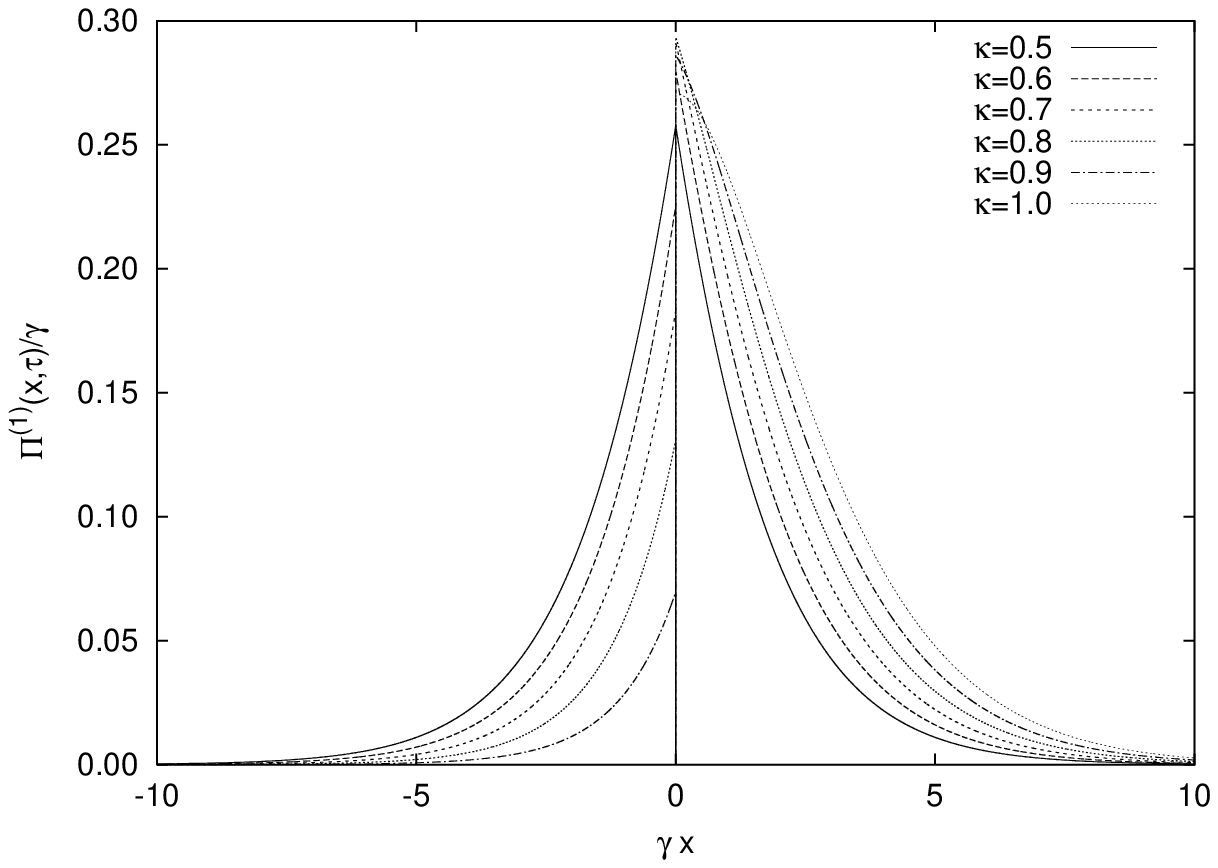}\\
\includegraphics[width=1\textwidth,keepaspectratio=true]{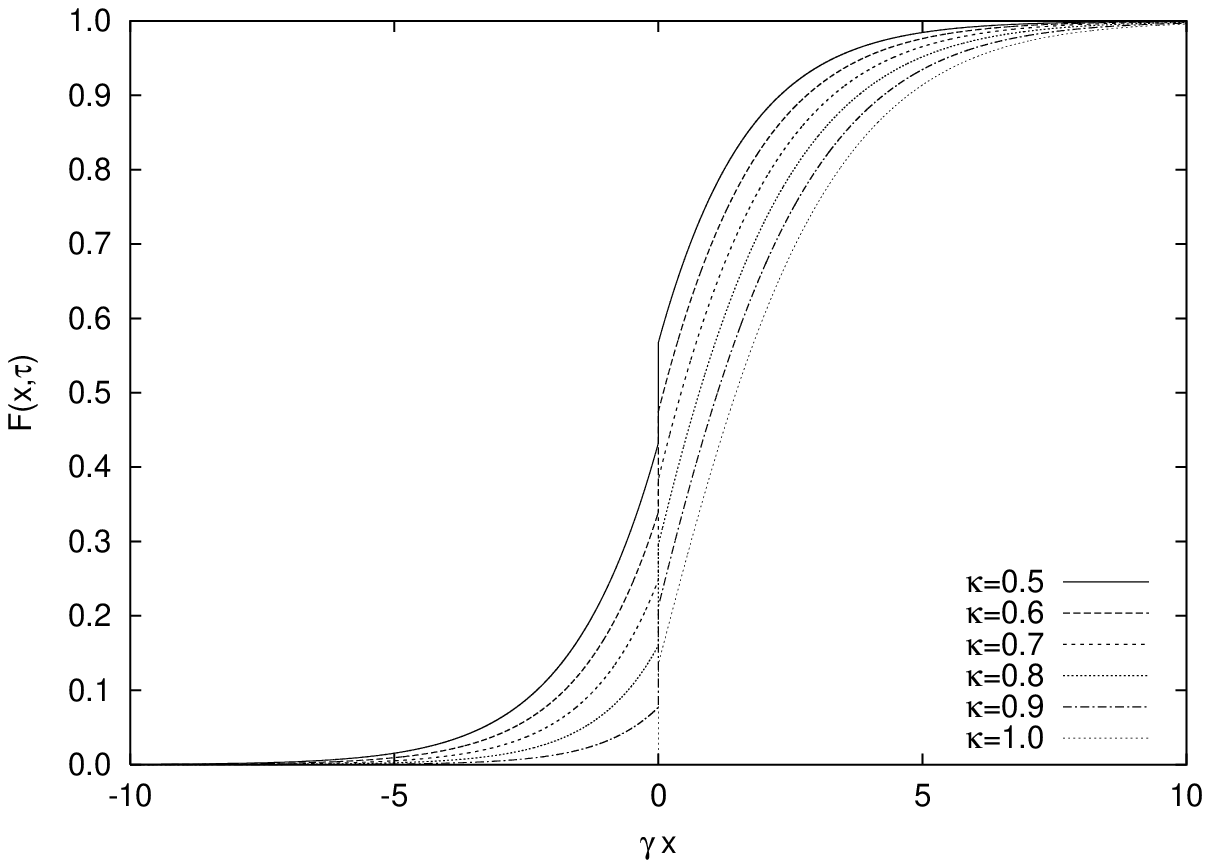}
\end{center}
\caption{Numerical inversion of the ``density contribution"  to
the propagator (upper panel) and the accumulated distribution
function (lower panel) for  exponential  waiting times with
intensity $\lambda$. The parameters were chosen such that
$\lambda \tau=2.0$. We consider several possible values of the
bias-related parameter $\kappa$, from the symmetric case
$\kappa=0 $ to the pure persistent case $\kappa= \f12$. Note the
discontinuity at $x=0$ in both plots.} \label{Fig2}
\end{figure}

For illustrative purposes, we consider  an example. Assume
  that  the
jump probability is exponentially distributed with parameter $\g$
but with different  probabilities to  jump  left and right:
\begin{equation} h(x)= \g  ( \frac 12+\kappa ) e^{ -\g x} \uu_{x\ge 0}+ \g
( \frac 12-\kappa ) e^{ \g x}
 \uu_{x <
0}, \label{h2}
\end{equation}
where   $ -\frac 12 \le  \kappa \le  \frac 12$ and $\g>0$ are
parameters.   In this case  we find $\hat{\Pi}^{(1)}(x ,s) =$
\begin{eqnarray}
 \frac{1-\hat{\psi}}{s}   \gamma
\hat{\psi} \Bigg\{\left[\frac{1-2 \kappa^2
\hat{\psi}}{\hat{\varphi} }+\kappa\right] e^{-[\hat{\varphi} +
2\kappa \hat{\psi}]\gamma x/2} \uu_{x > 0}
  +   \left[\frac{1-2\kappa^2 \hat{\psi}}{\hat{\varphi}
}-\kappa\right] e^{[\hat{\varphi} - 2\kappa\hat{\psi}]\gamma x/2}
\uu_{x < 0} \Bigg\}, \label{Pi_one}
\end{eqnarray}
where $\hat{\psi} \eq \hat{\psi}(s)$ and
 $\hat{\varphi}\eq
\hat{\varphi} (s)=2 \sqrt{ 1 -\hat{\psi}   +\kappa^2 \hat{\psi}^2
}
 $.
Inversion of the corresponding Laplace transform  can be
  accomplished when  $ \kappa = \frac 12$, $h(x)= \g    e^{ -\g x}\uu_{x \geq 0}$ and $\hat
  \psi(s)= \frac{\lambda}{\lambda+ s}$. In this case one finds that
 \beq  p(x,r+\tau|r)=\Pi(x ,\tau )=     e^{-\la   \tau}\delta (x )+   \sqrt{\frac{  \g  \lambda \tau}{x  }}
     e^{ -\lambda   \tau-\g   x}  I_1\big(2\sqrt{\lambda  \g   \tau
   x} \big)\mathbf{1}_{x>0},   \enq
where $I_n(\cdot)$ is the modified Bessel function of the first kind with order $n$.
For  general values of $\kappa$ one must resort  to  numerical
techniques.  In Fig.~\ref{Fig2} we perform a numerical inversion of
Eq.~(\ref{Pi_one}) corresponding to $\hat
  \psi(s)= \frac{\lambda}{\lambda+ s}$ and to different values of the parameter
  $\kappa$.   In the upper panel we plot the ``density contribution"  to  the propagator $\Pi^{(1)}(x,\tau)\eq
  \Pi(x,\tau) - e^{-\la   \tau}\delta (x) $ as a function of $\g x$
  for fixed value of time.
  Notice  how the discontinuity of    $h(x)$    at
  $x=0$  |cf. Eq.~(\ref{h2})| is inherited by $ \Pi^{(1)}(x,\tau)$. The lower panel  plots the accumulated distribution function
   $F(x,\tau)\eq \int_{-\infty}^x
  \Pi(x',\tau)dx'$, whose discontinuity comes from the delta contribution to $\Pi(x,\tau)$ at
  $x=0$. 

   Returning to the general case, we now derive the  propagator. Operating with the Fourier-Laplace transform and using
      (\ref{def_m}), (\ref{hat_phi}) and (\ref{hat_tilde_Pi}) we obtain that in the Fourier-Laplace
   domain, Eq. (\ref{prop_int}) reads
 \beq\hat{\tilde p}(\omega ,s|r)
=\big(1-\hat{\phi}(s|r)\big)/s+
  \hat{\phi}(s|r)
\tilde h(\omega)\hat{\tilde \Pi}(\omega ,s)  \label{p_ws} \enq
  The representation shows that the transition
probability contains   an    isolated  mass at $ x=0$ |which gives
the probability that no jump has taken place| and a sub-stochastic
density $ p^{(1)} (x, r+\tau|r)$. Upon inversion \beq
    p(x, r+\tau|r)= \d(x)\Big(1-\Phi(t|r) \Big)+ p^{(1)} (x, r+\tau|r), $$
    $$\text{where }  p^{(1)} (x, r+\tau|r)\eq \frac{1}{4\pi^2 i} \int_{c-i\infty}^{c+i\infty}
    \int_{-\infty}^{+\infty}e^{st -i\o x}
  \hat{\phi}(s|r)
\tilde h(\omega)\hat{\tilde \Pi}(\omega ,s)  d\o ds, \ c>0.
\label{prop}\enq
 Eq.~ (\ref{prop}) gives  the propagator
 in a fully explicit way. We skip a similar
expression for $\pi(x,\tau+r|r)$.

The above expressions simplify if $r\to\infty$, which  corresponds
to the steady-state solution. This limit is   relevant both from
the  mathematical and
   the   applied  point of view,
since it describes the case in which \it the only information available to the observer is the present value of the stochastic process\rm.
  Using equation
(\ref{p_ws}) and setting $\phi_\infty(\tau)\equiv \un{ r\to\infty}\lim \phi(\tau| r)$  the full   propagator reads
 \begin{eqnarray}
 \un{r\to\infty}\lim \hat{\tilde p}(\omega ,s|r)&=& \frac{1-\hat{ \phi}_\infty(s)}{s}+
\frac{\Big(1-\hat{\psi}(s)\Big)\hat{ \phi}_\infty(s)\tilde h(\omega)}{s
\Big(1-\hat{\psi}(s)\tilde h(\omega)\Big)} \label{formal}
 \end{eqnarray}
 Eq. (\ref{formal}) coincides formally with an
equation already reported in~\cite{JT76}. We admit that the
resemblance is not spurious, but it must be stressed that the
problem addressed there, although related, is different from
ours.  Nevertheless, we can go a  step further by evaluating $
\phi_\infty(\tau) $ explicitly; indeed,
 the renewal theorem yields that $\un{t\to\infty}\lim   m(t)   = t/\mu$, where $\mu\eq \int_0^{\infty} t\psi(t)dt$ is the mean sojourn
time. Inserting this into Eq. (\ref{hat_phi}) we obtain
 \begin{equation} \phi_\infty(\tau)
=  \frac{1-\Psi(\tau)}{\mu}, \quad \Phi_\infty(\tau)= \un{ r\to\infty}\lim \Phi(\tau| r)=1-\frac{1}{\mu}
\int_{0}^{\infty} t d\Psi(t+\tau).
\end{equation}

\section{Mean exit times}
\label{s4}   Another significant    magnitude in CTRW problems is
the mean exit time $T(x_r,r)$ off a given interval $A\eq [a,b]$.
This is defined as follows: suppose we observe that at  a certain
moment $r$ the process  was in
  $x_r: X(r)=x_r,\ x_r\in
  [a,b]$. Then $T_{[a,b]}(x_r,r)  $ is  the mean time that takes for the process  to exit   $A $. Note that for ease of notation we usually
   write $T_{[a,b]}(x_r,r)  \eq  T (x_r,r)  $.
 When $r$  is    one of the jump times,  $r=t_n$ say, we can simply write $\TT (x_n)\eq T_{[a,b]}(x_n,t_n)$.
    It can be  proven that $\TT(\cdot)$   satisfies the following  linear integral
  equation (see \cite{mmp05})

  \begin{equation}
\TT (x_n)= \mu+ \int_{a}^{b}h(x'-x_n)\TT (x')dx'. \label{METPi}
\end{equation}
We next show how to generalize these results     to obtain
$T_{[a,b]}(x_r,r) $. Let
 $  r+ \tau_r $ and  $ r+ t_{r }^{x_r}  $      be  the first jump time and, respectively, the first time past
 $r$  at which $X(t)$         exits  $
   A$ so    that $T_{[a,b]}(x_r,r) =< t_{r
    }^{x_r}>$|see section~\ref{s2} and  Fig.~\ref{Fig1}.
 Obviously $t_{r }^{x_r}  = \tau_r $   if   the process  exits $
A$ at
 the first jump after $r$    while  $t_{r }^{x_r} =\tau_r+ t_{  r+\tau_r  }^{
  y}$  if the process jumps to a position $y\eq X_{ r+\tau_r} $ \it still  within \rm $
 A$. Thus  we can  write
  $ t_{r }^{x_r}   = \tau_r+ t_{  r+\tau_r  }^{
 X_{ r+\tau_r}} \mathbf{1}_{X(r+\tau_r) \in  A}.  $
  By taking an average    we  find
 that
  \beq <  t_{  r+\tau_r  }^{
 X_{ r+\tau_r}} \mathbf{1}_{X(r+\tau_r) \in  A}> =  \int_a^b  < t_{  r+\tau_r }^{
 y} >  \Pr\big(X_{ r+\tau_r}\in dy|X_r=x_r\Big), \label{mean_t} \enq
  where, in the spirit of the total probability theorem,  we have conditioned on the after-jump position $X(r+\tau_r)$
   and used the ``lack of memory past jump-times" property, viz. Eq.~(\ref{memoryless}).  This expression is simplified further by noting that
       $< t_{  r+\tau_r }^{
 y} > =\TT (y)$. Further,   \beq\Pr\big(X_{ r+\tau_r}\in [y,y+dy]\big|X_r=x_r\big)=  \Pr\big(\Delta X \in  [y-x_r,y-x_r+dy]  \big)=h(y-x_r)dy, \enq
 since the jump amount is statistically independent of  the previous position. Hence, the  RHS of (\ref{mean_t})  equals $  \int_{a }^{b }\TT(y)
 h(y-x_r)dy$. By substitution we
 finally have
 \beq  T_{ }(x_r,r)=<\tau_{r }> -\mu+\TT(x_r)\eq \TT(x_r)+
\int_0^{\infty} t'\Big(  \phi(t'|r) -  \psi(t')\Big) dt'.\label{4.5}\enq

 Thus the mean  time to exit $[a,b]$ past $r$  is obtained by subtracting $\mu -<\tau_{r }> $ to the mean exit time after  a jump happened,
 $\TT(x_r)$ |which solves the   \it linear \rm integral equation~(\ref{METPi}).
 In section 5 (see Eq. (\ref{time}))  we evaluate this correction in the case  of Erlang  waiting
  times.

\section{Erlang sojourn times }
\label{s5}
 \subsection{Distribution and mean  of the
first jump past $r$}  \label{s5.1}

  Even
  although  for  a large  variety of    physical
  situations waiting
  times   can be fitted by an
exponential  distribution, there exist other possibilities  of
interest.  Here we consider an alternative distribution.
   Recall that a random variable is said to
have  Erlang distribution $\mathcal  Er( \nu,\lambda)$
    whenever   it can be expressed as a  sum of $\nu$, $\nu \in \N$,  iid
exponential  variables with parameter $\la$. It turns out that
$\mathcal  Er( \nu,\lambda)$ coincides with the
    Gamma
distribution $\Gamma( \be,\la)$ whenever $\be\eq \nu$ is an
integer.  This corresponds to \beq \psi(t)=\lambda \big(\lambda
t\big)^{\nu-1} {e^{-\lambda t}\over (\nu-1)!} ,\quad\hat
  \psi(s)=\Big({\lambda\over \lambda+ s}\Big)^\nu. \label{den} \enq
    To recover $\phi(\tau|r)$   we first consider the case of the renewal function, cf. Eq.~(\ref{def_m}).
 Here the integrand has poles at points $ b_j\equiv  \la(\e_j-1) $ where $ \e_j\equiv e^{{2\pi  ij\over \nu} }, j=1,\dots, \nu$.
  Hence, by closing  the contour appropriately  in the complex $s$-plane and
integrating  by residues  we find  that   the renewal function is
  (note that $b_{jR}\equiv \text { Re } b_j\le
 0$)

\beq  m'(t)=  \sum_{j=1}^\nu {\la\e_j \over
 \nu }  e^{  -\la (1- \e_j )t}   \  \text{  and  }  m (t)={\la t\over \nu}+ \sum_{j=1}^{\nu-1}
{ \e_j \over
 \nu (1-\e_j )}\big(1-  e^{  -\la (1- \e_j )t}  \big).\enq
Inserting this into   (\ref{hat_phi})   we find  upon
  Laplace inversion that
   \beq
\phi(\tau|r)=  {1\over 2\pi i} \int_{c-i\infty}^{c+i\infty} ds
e^{ s\tau } \sum_{j=1}^\nu {\la\e_j \over
 \nu(s-b_j)}  e^{   b_j   r} \Big[1-\Big({\lambda\over \lambda+
s}\Big)^\nu\Big]=     $$$$\sum_{n=1}^\nu \lambda \big(\lambda
\tau\big)^{n-1} {e^{-\lambda   \tau}\over
 (n-1)!}\a_n(r)  \  \text{ where }\a_n(r)= { 1 \over \nu}\sum_{j=1}^\nu  \e_j^n
e^{  b_j r}. \label{phi_E}  \enq  Further, the identity $
\sum_{j=1}^\nu  \e_j^n= \nu\d_{n\nu}$
   shows that  $\a_n(0)=\d_{n\nu}$ and $ \phi(\tau|0)=\psi(\tau)$.  Also
  \beq   \Phi(\tau|r)=1- \sum_{n=1}^\nu   \big(\lambda \tau\big)^{n-1}
{e^{-\lambda \tau}\over
 (n-1)!} \sum_{j=n}^\nu\a_{j}(r).\label{5.5}  \enq

We next evaluate the mean of this distribution. The upper
expression in  Eq.~(\ref{phi_E})  yields \beq \int_0^\infty
t'\phi(t'|r)dt'= - {d \hat \phi(s|r)\over ds}\Big|_{s=0}=
\Big({\nu+1\over 2 }+ \sum_{j=1}^{\nu-1}{ \e_j \over
  \e_j-1}  e^{   b_j   r}\Big)/\lambda.
\label{5.6} \enq In particular, if $\nu=2$,    then   \beq
\Phi(\tau|r)  = 1- e^{-\lambda \tau} \Bigg[ 1+  \Bigg({ 1+e^{-2
\lambda r }\over 2}\Bigg)\lambda \tau  \Bigg],\quad <\tau_r>\eq
\int_0^\infty t'\phi(t'|r)dt'=\frac{3+e^{-2 \lambda r
}}{2\lambda}. \enq

     Some of these calculations can be  extended to cover the
more general Gamma distribution $\Gamma( \nu,\la)$,  whose density
is still given by  Eq.~(\ref{den}), whenever $\nu$ is a rational
number: $\nu\equiv {n\over q}$ with $n,q$ irreducible integers.
Closing  the contour appropriately  in the complex $s$-plane   we
find   the derivative of  the renewal function  as \beq  m'(t)=
\sum_{j=1}^n {\la  \e_j \over
  \nu }  e^{    \hat b_j t}+{\la\over 2\pi i} \int_{-\infty}^0dxe^{\la t(x-1)}\Big({1\over e^{ 2\pi i  \nu }x^\nu-1}-{1\over  x^\nu-1}\Big).
  \label{m_prime}\enq
   In some cases the above integral can be evaluated in closed form. For instance, if $\nu =1/2$ we will have
\beq  m'(t)=  \sqrt{\frac{\lambda}{\pi t}} e^{-\lambda t}+
\lambda \text{Erfc}\big(-\sqrt{\lambda t}\big), \quad
m(t)=\sqrt{\frac{\lambda t}{\pi}} e^{-\lambda t}+ \frac{2\lambda
t+1}{2} \text{Erfc}\big(-\sqrt{\lambda t}\big)-\frac{1}{2},\enq
where $\text{Erfc}(\cdot)$ is the complementary error function.

\subsection{Propagator and escape times   under the  Erlang  distribution}
\label{s5.2}
   Given that     $ X_r=x$       the expected
remaining time to exit $[a,b]$, $a\leq x\leq b$,    follows by inserting
(\ref{5.6}) into (\ref{4.5}); since    the mean of Erlang distribution is
$\mu={\nu \over
   \la}$  we find that
    \beq  T_{ }(x_r,r)= \TT(x_r) + \Big({1-\nu \over 2\la}  +
\sum_{j=1}^{\nu-1} { \e_j \over
    \lambda(\e_j-1)}  e^{   b_j   r} \Big)\label{time}. \enq
  In particular, if $\nu=2$,    then  $  T_{[a,b]}(x_r,r)= \TT(x_r)
 -\Big(1- e^{-2\lambda  r}  \Big)/2\la $.

 We next consider the  after-jump transition
probability. Note that for Erlang times $\mathcal Er(\nu,\lambda)$
Eq.~(\ref{hat_tilde_Pi}) yields
 \beq  \tilde \Pi (\omega ,t)={1\over 2\pi i}
\int_{c-i\infty}^{c+i\infty} {ds\over s} e^{ st }\frac{
(\la+s)^\nu-\la^\nu}{{  (\la+s)^\nu-\la^\nu \tilde h(\omega)  }}.
\enq The integrand has poles   at   $ s_j=\la \big( q(\o)\e_j-1),
j=1,\dots,\nu$ where $q(\o)$ is a determination of $  \tilde
h^{1/\nu}(\o)$. Note that $0\le | \tilde h | \le 1$ and \it all
poles
   are located on the left
half-plane\rm. Thus, Cauchy's
 theorem yields that the ``after-jump" propagator is  $
   \Pi(x,t)= \d(x)\Big(1-\Psi(t) \Big)+\Pi^{(1)}(x,t)$, where
   \beq\Pi^{(1)}(x,t) ={1\over 2\pi  }  \int_{ - \infty }^{
\infty } e^{-i\omega x -\lambda t}      \sum_{n=1}^{\nu} \Bigg( {
\e_n \over
 \nu q^{ \nu-1}}{ \tilde{h}-1\over     \e_nq-1}  e^{   \la\e_n qt}-{\big(\lambda t\big)^{n-1}\over (n-1)!}\Bigg)d\o. \label{5.11}
  \enq

  When $\nu=1$, (\ref{5.11}) is the well-known inversion formula for a
  CPP.
When $\nu=2$  one has that  \beq   \Pi^{(1)}(x,t)  = {1\over 2\pi
 }  \int_{ - \infty }^{   \infty } e^{-i\omega x -\lambda
t}    \Bigg[{1\over \sqrt {\tilde h}} \sinh \Big( \la  \sqrt
{\tilde h }  t\Big)+\cosh\Big(\la \sqrt {\tilde h }  t\Big)-1-\la
t\Bigg] d\o. \label{49}\enq
  Note that, $\Pi^{(1)}(x,t)=O(t^2)$ as $t\to 0$,  which confirms again that the
  process can not be Markov.

\begin{figure}[hbtp]
\begin{center}
\includegraphics[width=1.0\textwidth,keepaspectratio=true]{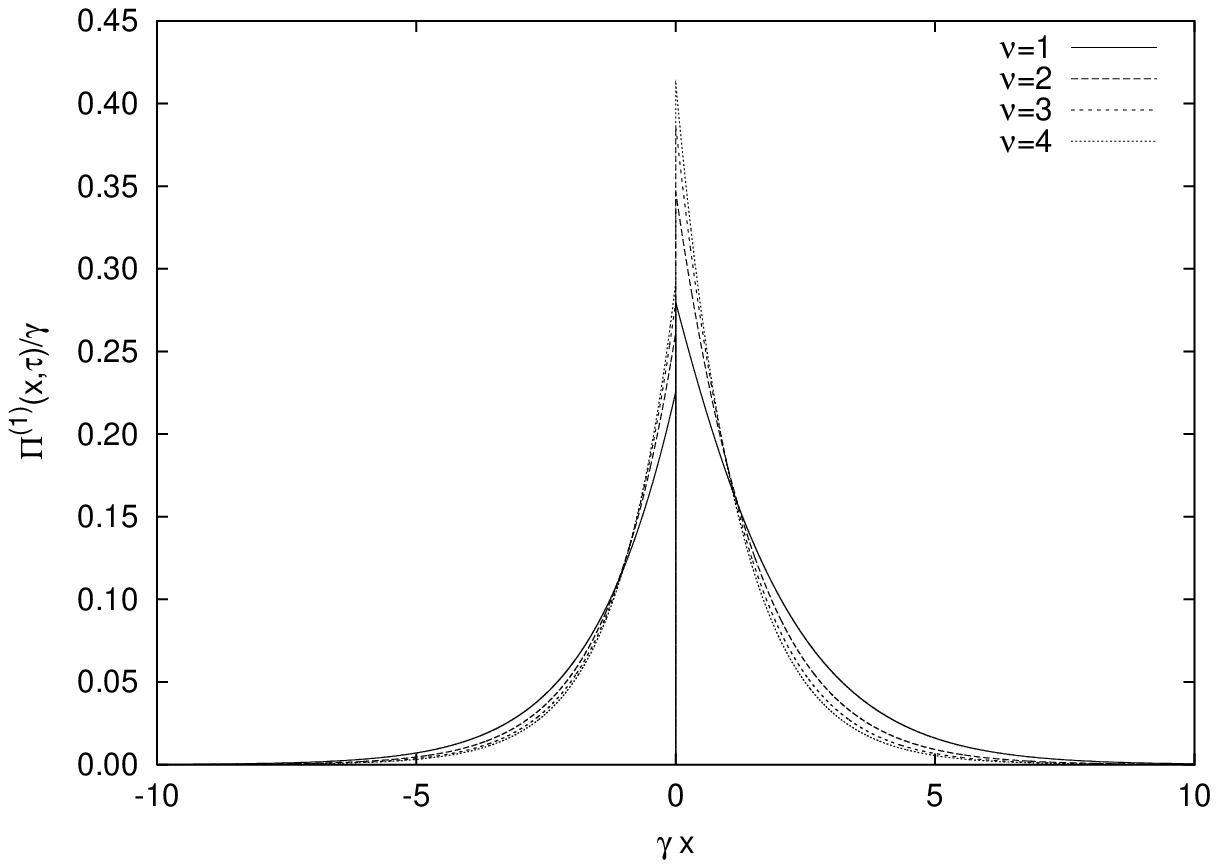}
\includegraphics[width=1.0\textwidth,keepaspectratio=true]{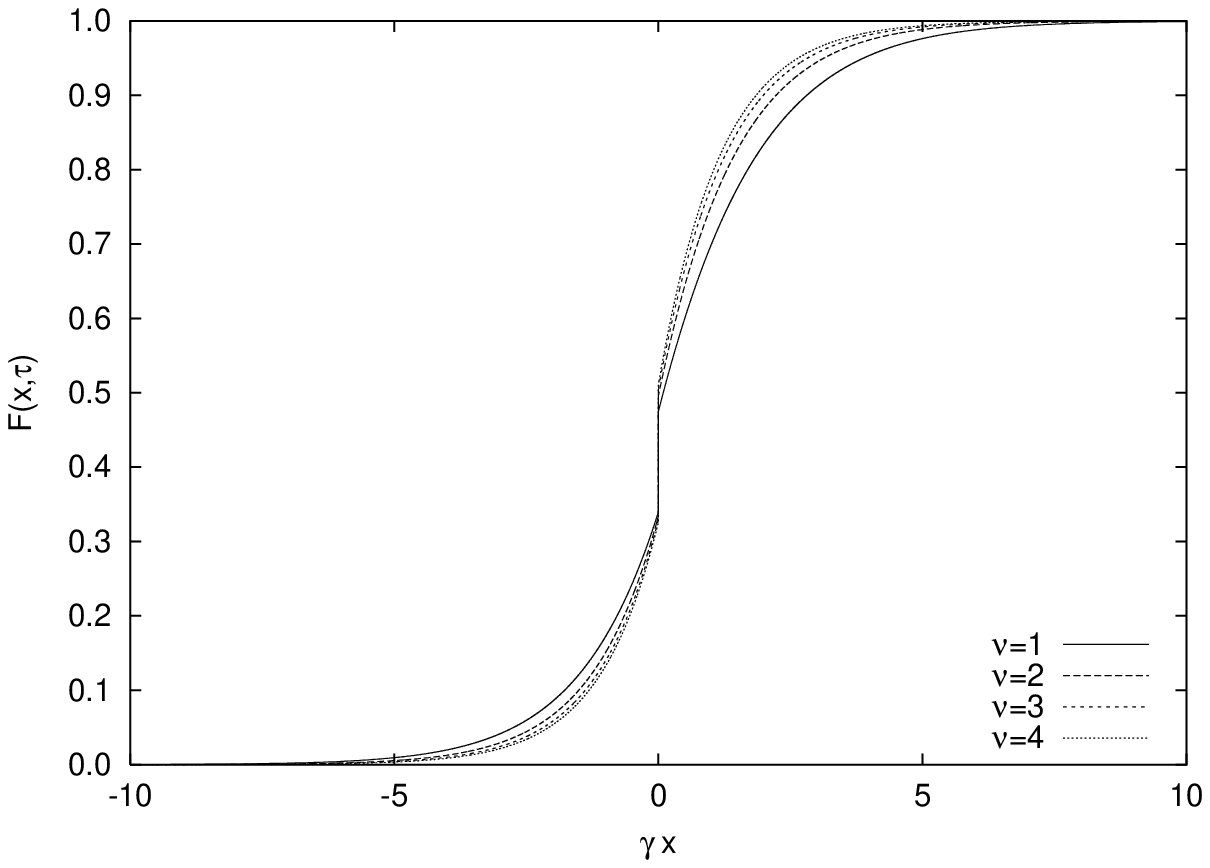}
\end{center}
\caption{Numerical inversion of the ``density contribution"  to  the propagator (upper panel) and the accumulated distribution function
 (lower panel) for Erlang waiting times.  The intensity $\lambda$ is different for different values of $\nu$: $\lambda =2.00 \tau^{-1}$ ($\nu=1$),
 $\lambda \approx 3.27 \tau^{-1}$ ($\nu=2$), $\lambda \approx 4.48 \tau^{-1}$ ($\nu=3$), $\lambda \approx 5.65 \tau^{-1}$ ($\nu=4$).
 This choice ensures $1-\Psi(\tau)=e^{-2}\approx 0.14$.} \label{Fig3}
\end{figure}

 In Fig.~\ref{Fig3} we perform a numerical inversion of
Eq.~(\ref{5.11})   corresponding to the jump density $h(x)$  of
Eq.~(\ref{h2}) where  $\kappa=0.1$ and  we take different values
of the parameter $\nu$, concretely $\nu=1,2,3,4$. In order to keep
the plots commensurable we have changed the $\lambda$ parameter in
such a way that the no-jump probability $1-\Psi(t)$ remains fixed.
Observe the different decay behavior which is noticeable both in
$\Pi^{(1)}(x,\tau)$  and in the accumulated distribution function
$F(x,\tau)$.

The   propagator $ p(x,\tau+r|r)$  can be evaluated in closed
form when  $r\to\infty$. For  $\nu=2$, say,  upon inversion of
(\ref{formal}) with (32), we find that $p_\infty(x,\tau )\equiv
\un{r\to\infty}\lim p(x,\tau+r|r)  $ reads \beq p_\infty(x,\tau )=
\Big(1-\Phi_{\infty}(\tau)\Big) \delta(x)+ {1\over 2\pi
 }\int_{ - \infty }^{  \infty } e^{-i\omega x -\lambda
\tau}    \Bigg[\frac{1+\tilde{h}}{2 \sqrt{\tilde h}} \sinh \Big(
\lambda  \sqrt {\tilde h }  \tau \Big)+\cosh\Big(\lambda \sqrt
{\tilde h }  \tau\Big)-1-\frac{\lambda}{2}  \tau \Bigg] d\omega.
\label{pinf}
 \enq
Notice how it does not quite settle to the after-jump propagator
$\Pi(x,t)$ of Eq. (\ref{49}). In particular for $\tau\to 0$,
$p_\infty(x,\tau )$  exhibits a  quite  different behavior, since
 Eq.  (\ref{pinf}) implies that
 \beq p_\infty(x,\tau )-\Bigg(1-\frac{\tau}{\mu}\Bigg) \delta(x)=  \frac{h(x)}{\mu} \tau +O(\tau^2),\label{52}
 \enq where we recall that $\mu=2/\lambda$.  We point out  that  the expansion    (\ref{52})  also holds  in  the  general case as can
 be  easily seen  using  Eq.  (\ref{formal}).

\ack

The authors acknowledge support from MEC under contracts No.
SCO2006-28541-E (JV and MM), FIS2006-05204-E (MM) and
FIS2005-01375 (JV) and by
  Junta  de Castilla-Le\'on   VA0135C05 (JV).

\end{document}